\documentclass{llncs}

\usepackage{hyperref}
\usepackage{cite}
\usepackage{graphicx}
\usepackage{mathrsfs}
\usepackage{algorithm}
\usepackage[noend]{algorithmic}
\usepackage{paralist}

\usepackage{listings}
\usepackage{placeins}

\usepackage{amssymb}
\usepackage{xcolor}
\usepackage{titling}

\definecolor{airforceblue}{rgb}{0.36, 0.54, 0.66}
\definecolor{cadmiumgreen}{rgb}{0.0, 0.42, 0.24}
\definecolor{blue-violet}{rgb}{0.54, 0.17, 0.89}
\definecolor{blue(ryb)}{rgb}{0.01, 0.28, 1.0}
\definecolor{bostonuniversityred}{rgb}{0.8, 0.0, 0.0}
\definecolor{burntorange}{rgb}{0.8, 0.33, 0.0}

\definecolor{grey}{rgb}{0.4,0.4,0.4}
\definecolor{black}{rgb}{0.0,0.0,0.0}
\definecolor{blue}{cmyk}{1,0,0,0}
\begin{document}

\title{Integrating Inter-Object Scenarios with Intra-object Statecharts for Developing Reactive Systems} 
\vspace{0.5cm}

\titlerunning{}

\author{David Harel \and Rami Marelly  \and Assaf Marron \and Smadar Szekely}
\authorrunning{Harel et al}

\institute{The Weizmann Institute of Science, Rehovot, Israel
\\\{david.harel,  rami.marelly, assaf.marron, smadar.szekel\}@weizmann.ac.il}

\maketitle

\begin{abstract}
In all software development projects, engineers face the challenge of translating the requirements layer into a design layer, then into an implementation-code layer, and then validating the correctness of the result. Many methodologies, languages and tools exist for facilitating the process, including multiple back-and-forth  `refinement trips' across the requirements, design and implementation layers, by focusing on formalizing the artifacts involved and on automating a variety of tasks throughout. 
In this paper, we introduce a novel and unique development environment, which integrates scenario-based programming (SBP) via the LSC language and the object-oriented, visual Statecharts formalism, for the development of reactive systems. LSC targets creation of  models and systems directly from requirement specifications, and Statecharts is used mainly for specifying final component behavior.  
Our integration enables semantically-rich joint execution, with the sharing and interfacing of objects and events, and can be used for creating and then gradually enhancing testable models 
from early in requirements elicitation through detailed design. In some cases, it can be used for generating final system code. 
We describe the technical details of the integration and its semantics and discuss its significance for future development methodologies.
\end{abstract}
\section{Introduction}

When developing software in any discipline, using the traditional waterfall process or any variant of agile and spiral development, all stakeholders are faced with the existence of multiple conceptual layers: requirements, design and final running code.  
Throughout the development process, domain experts, system engineers, programmers and other stakeholders constantly interact to make sure that the transitions across the boundaries of such conceptual layers are indeed correct, and offer an acceptable mapping, usually one of uni-directional refinement. 
Development tools assist in the process, by introducing artifacts that can be understood and discussed by  people of different professional backgrounds, and which can be tested and validated, manually or automatically, against artifacts from a different layer. 

More specifically, functional requirements describe system behavior and traits, from
the point of view of the various stakeholders.  
They often consist
of scenario-based descriptions of sequences of events that reflect
desired, allowed and forbidden behavior. A central characteristic of such scenario-based
specifications is their inter-object nature. A scenario can contain
a flow of events involving any number of objects, 
internal or external, including subsystems and human users,  
for example, in the Windows operating system \emph{``when the user presses \texttt{ctl} and then \texttt{alt} and then \texttt{del}, and does not release the pressed buttons, then the task manager screen is displayed''}. Each scenario can interweave listening out for many events, triggering any number of actions, and subjecting all operations to a variety of conditions. 
Each requirement scenario is self standing, and with sufficient context can appear anywhere  in a requirements document. The composition and dependencies are well understood in the reader's mind because of the intuitiveness of the compositional idiom.  

In addition to natural-language descriptions in requirement
documents, such scenarios are often expressed in rigorous
 languages. A good example is the  visual
language of \emph{Live Sequence Charts} (LSCs)~\cite{DH01a,hm03}, which evolved from Message
Sequence Charts (MSCs).
The LSC concepts were adopted
in the later formalization of UML sequence diagrams  and in a variety of tools and methodologies. Detailed semantic definitions have made it possible to
simulate and execute  these scenario-based specifications via run-time concurrent consideration of all scenario constraints and preferences (a process termed \emph{play-out}).
This gave rise to the inter-object paradigm of {\it scenario-based programming} (SBP), also termed \emph{behavioral programming}, originally supported by the
Play-Engine~\cite{hm03} and later by  PlayGo~\cite{playgo}.
SBP was later extended to standard programming languages like Java, C++, JavaScript and Erlang (see, e.g., ~\cite{bpcacm}),  and to domain-specific textual modeling languages like ScenarioTools's SML~\cite{Greenyer2016a}. 

While inter-object scenarios are an excellent way to specify and compose requirements, in the common approaches to system design and
implementation, system behavior is constructed from {\emph intra}-object specifications. These object-oriented (or object-centric) descriptions provide for each object separately 
its  behavior as manifested in direct interaction with the environment and with other objects, through events, message exchanges and  internal operations. 
There are numerous non-visual procedural languages for object-oriented
programming, such as Java and C++, but for a formal visual description of
reactive behavior, it is common to use state machines,
where each describes all the states of a given object and its reactions, in each state, to all possible
external and internal stimuli.

In 1987, the \emph{Statecharts} language~\cite{SC_DH87} was introduced, as a visual
formalism that augments conventional state machines with
notation and semantic definitions for the concurrency and hierarchy
necessary to specify and then directly execute complex behavior. An object-oriented version thereof was described in~\cite{HG97}, and among other things this variant became the basis for the state-based language of the UML.
Statecharts have been
implemented in multiple software engineering tools, such as
STATEMATE  and Rhapsody (acquired by IBM),
MATLAB's Stateflow, SCADE, LabView,
Yakindu~\cite{yakindu}, and others, 
and have become the visual formalism of choice for
intra-object behavior specification in a multitude of industries.

\begin{figure}[htb]
\centering
\includegraphics[scale=0.4]{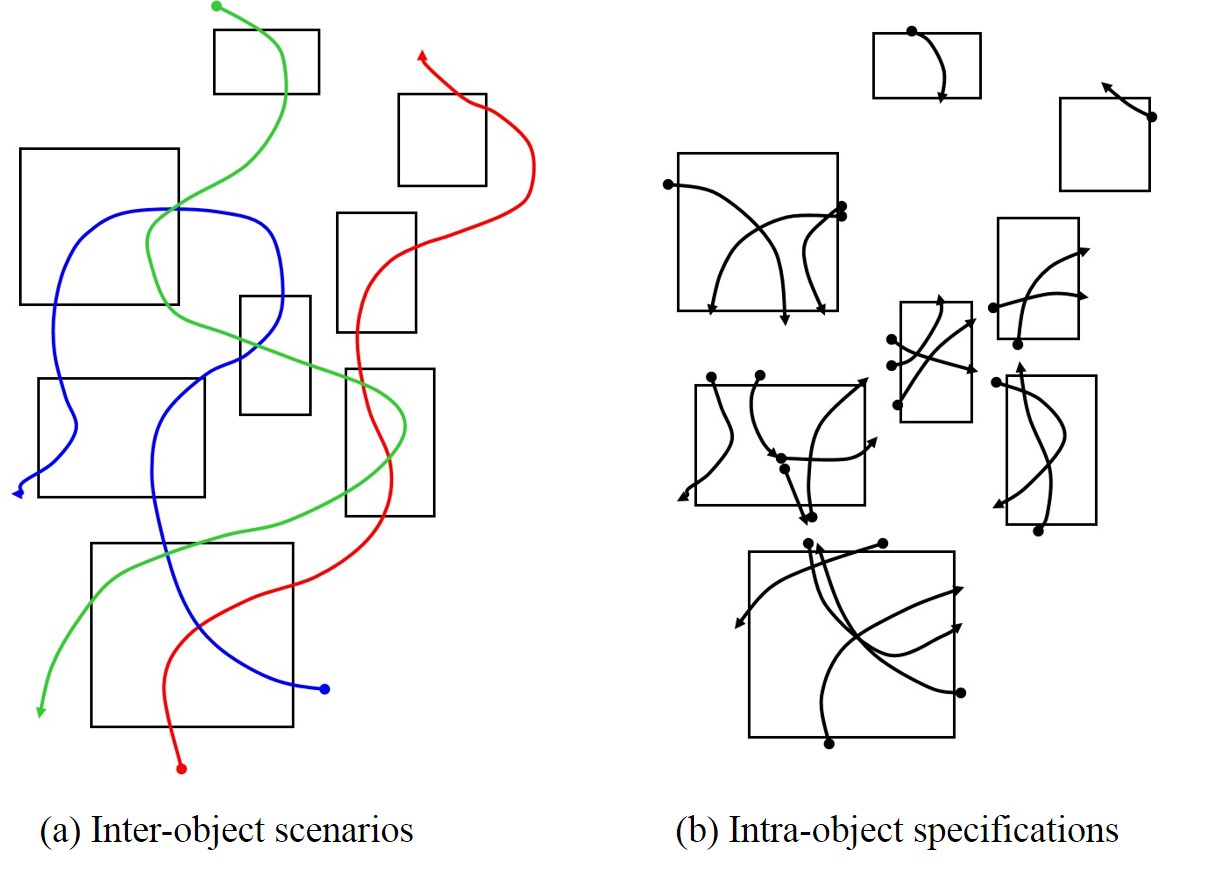}
\caption{{\footnotesize From \cite{hm03}: (a) Inter-object scenarios cross multiple object boundaries in describing `full stories';  (b) intra-object specifications describe the `full reactivity' of each object.} }
\label{snakesinterintra}
\end{figure}

The conceptual duality between the inter-object and intra-object
approaches is illustrated in Fig.~\ref{snakesinterintra}, originally
appearing in \cite{hm03}.

The `full story' of the sequence of events in each scenario is provided
explicitly in the inter-object scenarios, while it is only implicit in
the intra-object specification. Conversely, the full reactive behavior of
any given object is visible explicitly in the intra-object
specification, but in scenario-based specifications it must be derived from multiple scenarios.

While the intra-object specification approach is directly aligned with
classical object-oriented programming, the translation from an inter-object, scenario-based specification
to implementation is a central issue in software engineering, and constitutes a substantial part of many software development efforts.

In the past, scenario-based behavior specifications were used mainly to help guide
the development, and then the testing, of the implementation --- given in a conventional intra-object fashion. 
Testing is often done with the aid of a tool that  monitors the execution of the intra-object implementation and confirms that the specified inter-object scenarios  indeed comply with those specific runs. 
A key contribution of  SBP is the fact that the LSC language and its derivatives have powerful enough syntax and semantics as to render the requirement specifications directly executable. In other words,  SBP enables building a working system (or a highly functional simulator thereof) from modules that are aligned with how humans often describe behavior.  
What happens during the running, or playing out, of the specification  is that an SBP execution engine follows all the
scenarios in parallel, waits for environment and system-driven events and reacts to them by
triggering other events according to the specified behavior, while, significantly, 
avoiding or delaying the triggering of events that are forbidden (in the current state) by some scenario.

This allows for direct execution and dynamic testing of requirements in early prototypes and simulations, 
and/or for programming a system using its multi-modal requirement scenarios (see,e.g., in list~\cite{DHpubs}, paper 174). This can save the developers
 and engineers part of the efforts associated with transforming requirements into design. 
Solutions for key design considerations (or concerns...), such as detecting conflicts between independently-specified scenarios, or efficient parallel execution of thousands of scenarios, are emerging from research on
 SBP (see, e.g.,~\cite{bpcacm} and, in list~\cite{DHpubs} papers 230, 233, 257).

SBP supports agile, or spiral, development methodologies, in that when new requirements or refinements are introduced, one can often
specify them incrementally in new stand-alone scenario with little or no change to existing ones (see~ \cite{DHpubs} paper 230). 
The naturalness of programming with scenarios was further discussed in several studies and in observing how children learn to program.

However,  SBP has its limitations. While  early in  development  external system behavior is usually conveniently  described using scenarios, there are many inner mechanisms and details that are less amenable to such specification and require an object-oriented method. Together with constraints of pre-existing OO software components, and ingrained programming tradition, this often causes developers to make the entire design intra-object.

\begin{figure} [htb]
\centering
\includegraphics[scale=0.3]{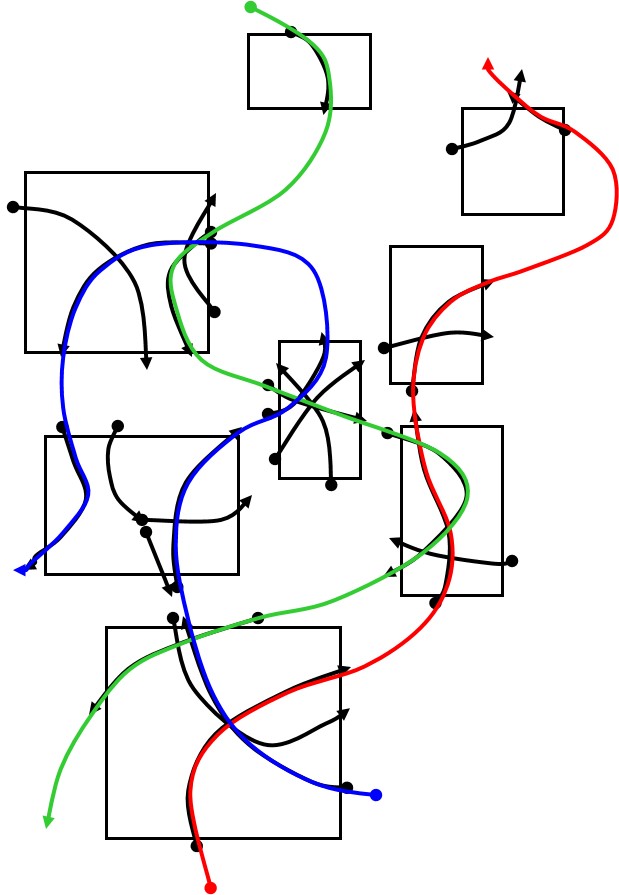}
\caption{{\small A modest illustration of our vision: the inter-object and intra-object views of system behavior are cohesively integrated, both superimposed upon each other and complementing each other.}}
\label{combinedInterIntra}
\end{figure}

In this paper, we present an overall development philosophy, which supports a natural integration of  inter-object and
intra-object methodologies. It offers a gradual and coherent transition from the former to the latter, allowing the coexistence and co-execution of `completed' intra-object Statecharts with inter-object scenarios that  have  not yet been implemented in Statecharts,  or which have been purposely retained; e.g., for verification and validation. (see Fig.~\ref{combinedInterIntra}).

Specifically, we have extended the PlayGo  tool for LSCs and have integrated it with the Yakindu Statechart 
tool.
The integrated  tool supports beginning with an executable model of the requirements and
incrementally adding  implementation details by object-oriented
Statecharts, and then optionally removing already-implemented requirements.
Thus, the proposed approach and tool support the smooth back-and-forth transition across boundaries of the conceptual layers of requirements elicitation, formal specifications design and implementation.

In Section \ref{StatechartsLSCs}, we give a brief overview of the Statecharts and LSC languages (more details can be found in \cite{HG97, hm03}, respectively). In  Section \ref{SimpleExample}, we demonstrate  the principles of the new development approach through an extremely simple example where we transition from a pure scenario-based specification to an implementation that contains Statecharts and, optionally, LSCs. In Section\ref{Implementation}, we briefly review the details of the semantics and mechanics of the integration, dealing with issues such as event selection, concurrent execution,  object mapping and sharing and more. Further technical details, including rigorous semantic definitions, are available in the supplementary material at \url{www.b-prog.org/sctlscSupp}~. In Section \ref{Railcars}, we  show more of the capabilities of the solution and some of the finer points of the semantics, via a more elaborate case study, which revisits the railcar system, first introduced with object-oriented Statecharts in~\cite{HG97}. In Section \ref{RelatedWork}, we review earlier efforts in this area wrt the present contribution, especially in light of the  tightly-coupled joint execution and shared object model that we introduce. We conclude in Section \ref{Conclusion}.  

\section{Introducing the Statecharts and LSC languages} \label{StatechartsLSCs}

\subsection{Statecharts}

\begin{figure} [h!]
\centering
\includegraphics[scale=0.7]{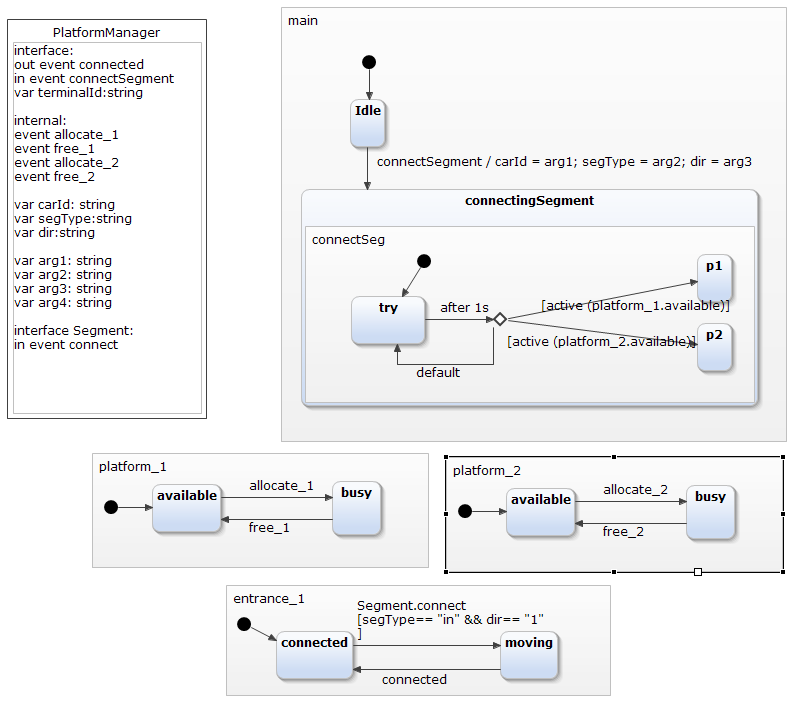}
\caption{A statechart of the platform-manager object of the railcar example, showing parts of its behavior, in four concurrent states.
 The two station platforms, {\it Platform\_1} and {\it Platform\_2,} can be allocated (or freed) for an incoming  (resp., departing) railcar. The {\it Entrance\_1} state represents the current status of the rail segment that connects  {\it Entrance\_1} with the platforms. The superstate  {\it main} moves from {\it Idle} to {\it connectingSegment} upon the triggering of the {\it connectSegment} event, 
in which case its three arguments, {\it arg1}, {\it arg2} and {\it arg3}, are stored in the internal variables {\it  carID}, {\it segType} and {\it dir}, respectively. When entering the {\it connectingSegment} superstate, the platform manager tries to allocate a platform to the incoming railcar, by checking which platform is available, in intervals of one second, until successful. This is done using the {\it choice} construct and the {\it active} function checks if another region is at a certain (sub-)state. 
}
\label{statechartexample}
\end{figure}

Three key concepts that the Statecharts formalism added to classical state machines are (i)~concurrency, i.e., separate states that are active and can make transitions at the same time as other states;  (ii)~hierarchy, i.e., the ability to specify that one state contains multiple other states and associated transitions, with an unbounded containment depth; and, (iii)~the ability to condition a local behavior on the fact that another region is in a particular (sub-)state .
States are drawn as rountangles (rectangles with rounded corners).
Concurrent regions, also termed orthogonal, are drawn either as separate, free-floating rectangles  when at the top level, or as partition with vertical dashed lines of states. 
Hierarchical containment is depicted in the physical containment of a Statechart diagram within a region within a state. See example in Fig.~\ref{statechartexample} (taken from the railcar application described in Section~\ref{Implementation}).  

Statechart transition arrows can be (optionally) labeled with (i) events that trigger the transition; (ii) a guard condition (in square brackets) that must be true to enable it; and (iii) actions that are to be carried out when the transition is taken (specified following a `/').  Additional actions can be specified to be taken upon entering or exiting a state, or while in a state.

The Statecharts language
contains additional features (see also Fig.~\ref{statechartexample}), including specifying the raising or triggering of events; richer specification of conditions and time,    
ability to re-enter a super-state directly to the inner state in which  it was when the super-state was previously exited,  dealing with synchrony and parallelism/simultaneity (like the ordering of events that become enabled `at the same time'),  reference to other objects and states within the statecharts of  those objects, and more.

In the Yakindu Statechart tool, which we use in our implementation, every statechart specification contains also a list of interfaces representing the class to which this statechart belongs, and objects or classes with which the statechart can communicate and the related events and variables. 

A key contribution of our integration is that the object model used by LSC is the very same one that is used by the Statecharts infrastructure.

\subsection{Live Sequence Charts (LSC)}
\begin{figure} [h!]\label{LSCsExample}
\centering
\includegraphics[scale=0.5]{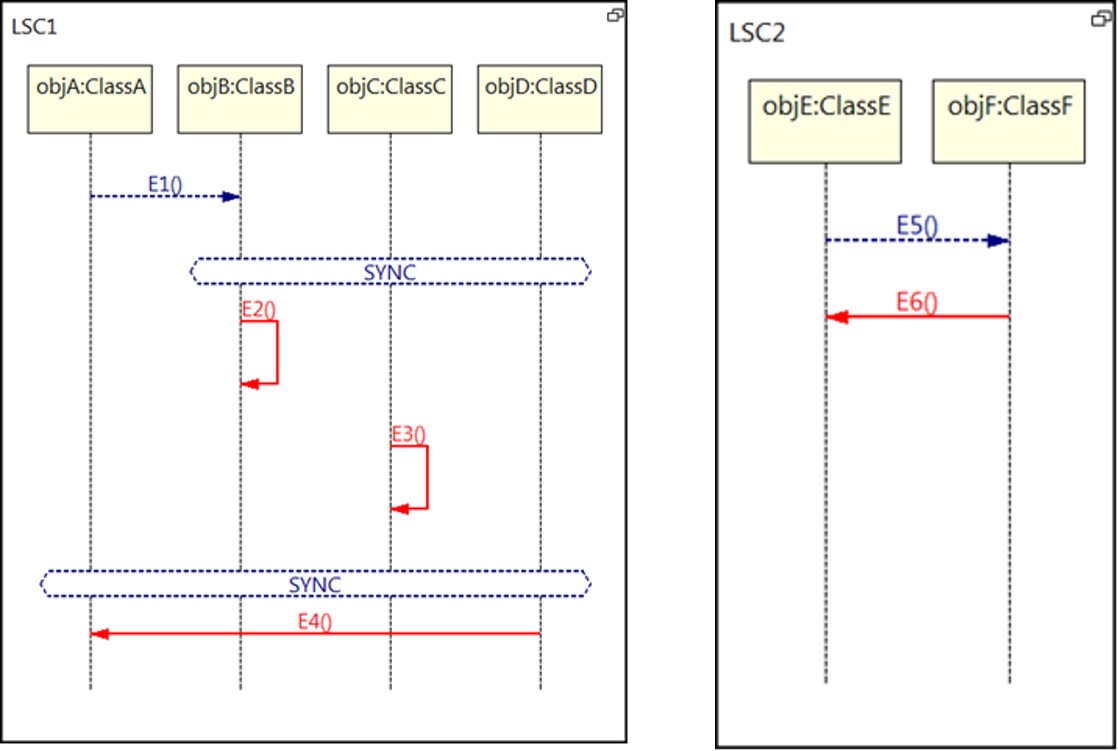}
\includegraphics[scale=0.5]{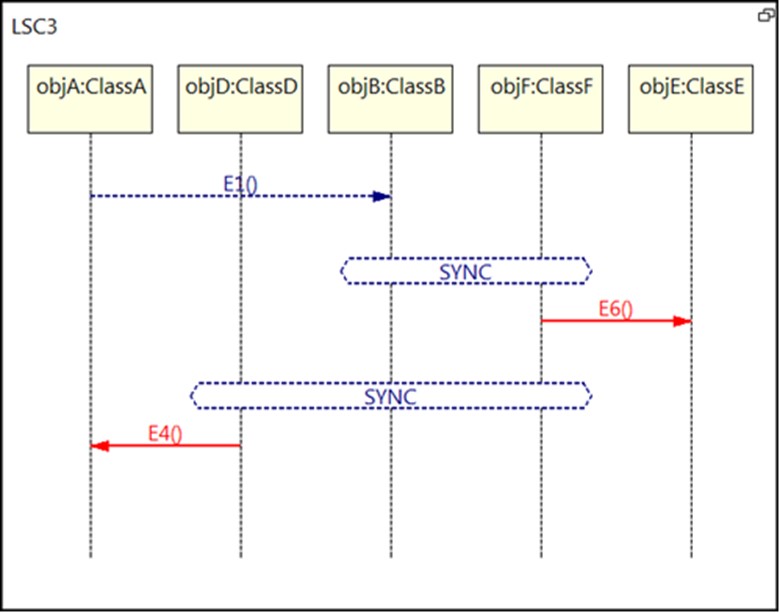}
\caption{Live Sequence Charts (LSCs) example: 
Scenario LSC1 specifies that after event E1 occurs, events E2 and E3
must occur, in any order, and, after both of them occur (enforced by the SYNC construct), E4 must
occur. LSC2 specifies that after E5 occurs E6 must
occur, and LSC3 specifies that once E1 occurs, E4
cannot occur until E6 occurs. Hence, when executing these 
LSCs, after E1 is triggered E4 will be delayed until E5 is triggered (by the environment or by some other scenario), subsequently triggering E6 and enabling E4.
}
\end{figure}

Fig.~\ref{LSCsExample}
shows several LSCs (The acronym \emph{LSC} is both the language name  and a noun for a single scenario (\emph{plural: LSCs})).  
Each scenario describes one aspect of system behavior --- typically its  response to an event or a sequence of events under certain conditions. The events are messages (depicted as horizontal  arrows) exchanged between (vertical) lifelines. Each lifeline is associated  (labeled) with an object (symbolically by class, or concretely using a particular instance thereof).
In a given lifeline, events are ordered, with time flowing from
top to bottom. 
The order among events that appear on different lifelines is partial and can be constrained  by other language constructs that synchronize those lifelines.

The LSC language distinguishes between events that are executed, i.e., triggered by the run-time infrastructure when enabled (marked by solid lines), and events that are merely monitored, i.e., waited for, in the particular scenario (marked by dashed lines). 
The  language also distinguishes events that once enabled, \emph{must} eventually 
occur (colored red), from events that only \emph{may} occur (colored blue). 

The LSC language supports additional constructs, such as conditions, including ones that can cause interrupts and breaks in scenarios, variable assignment, flow control (e.g. loops), nesting of subcharts, and more.
 
The PlayGo development environment for LSC provides a rich GUI for class/object model specification, scenario specification --- both by drawing and by using natural language (English text), execution (play-out), including simulation and debugging, and play-in (translating series' of GUI-based user-controlled event triggering into scenarios). 

%

\section{Integrating LSCs and Statecharts: A Simple Example} \label{SimpleExample}

In this section, we illustrate the integration of scenarios and statecharts (and also discuss some of the motivation for it), using an extremely simple example. We focus only on the value in terms of the development process that  integrates the two models,  and do not deal with problem-specific nuances. A more elaborate example is given later, in Section\ref{Railcars}.

\begin{figure} [h!]
\centering
\includegraphics[scale=0.4]{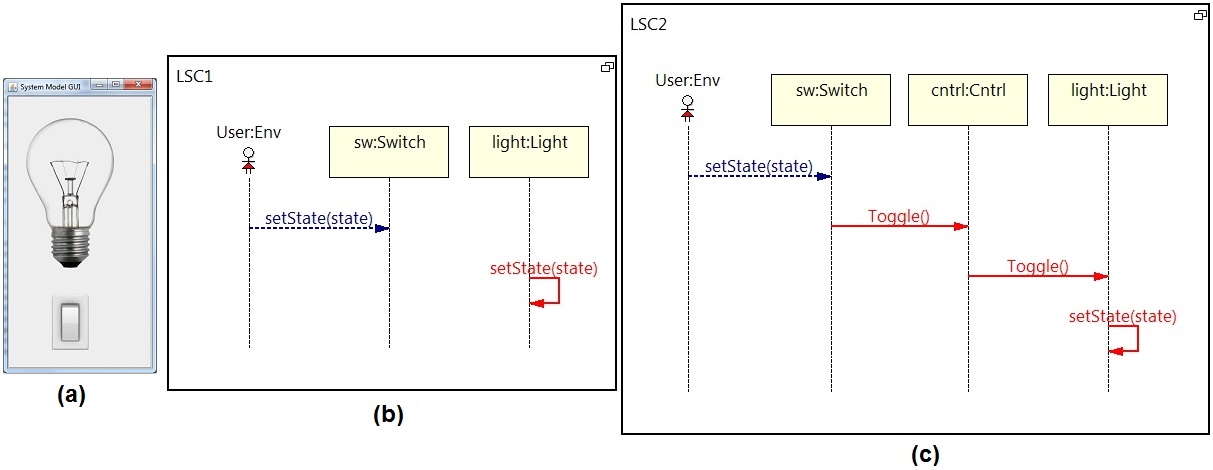}
\caption{The Switch-Light system: requirements and design}
\label{SL1abc}
\end{figure}

The system is composed of a simple GUI of a switch and a light  shown in  Fig.~\ref{SL1abc}(a).

The only requirement is that whenever the switch is set to on the light turns on, and when it is set to off, the light turns off. This is is coded as a single scenario, shown in Fig.~\ref{SL1abc}(b). 

Note the use of the same variable name, {\it state}, and values  {\it on} and {\it off} for the switch and the light, to create intuitive scenario logic. In this phase, the user does not care or know how the system implements the requirement; e.g.,  how the information about the switch's state is transferred to the light.

This LSC is executable, and using PlayGo the user can already test the specification by turning the switch on and off via the GUI and checking the light's reaction.
In the next step, the developer starts to incorporate design considerations, by introducing a controller. The controller receives a {\it toggle} message from the switch and sends a {\it toggle} message to the light. This LSC (shown in Fig.~\ref{SL1abc}.c) is executable as well, and while running it the developer can track the sequence of events between the switch, the controller and the light.

\begin{figure} [h!]
\centering
\includegraphics[scale=0.45]{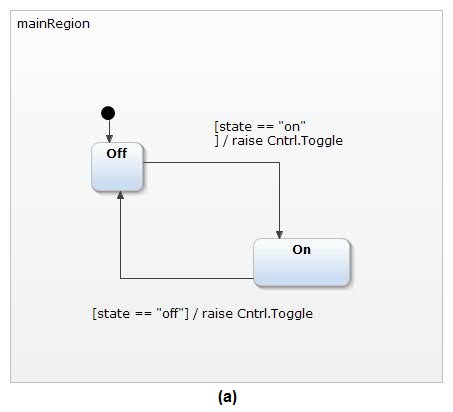}
\includegraphics[scale=0.45]{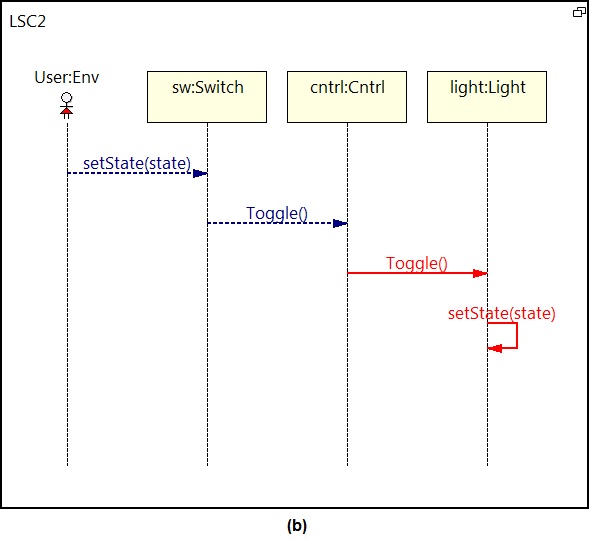}
\caption{The Switch-Light system: implementing the switch}
\label{SL2ab}
\end{figure}

Now that the design is completed, the developer moves to the implementation phase. The first thing they like to do is to implement the switch's logic, which can be done by the statechart of Fig.~\ref{SL2ab}.(a). This  moves the responsibility for the switch's behavior from the LSCs to the statechart, and the corresponding message in the LSC is changed from {\it executed} to {\it monitored}  (shown in Fig.~\ref{SL2ab}.(b).

This integrated model, consisting of an LSC and a statechart, is also executable: When the user clicks the switch in the GUI,
the statechart reacts to the event by calling the {\it toggle} method of the controller. This event is ``caught" by PlayGo and is unified with the monitored message, thus allowing PlayGo to proceed, executing the next {\it toggle} message, and then turning the light on.

\begin{figure} [h!]
\centering
\includegraphics[scale=0.45]{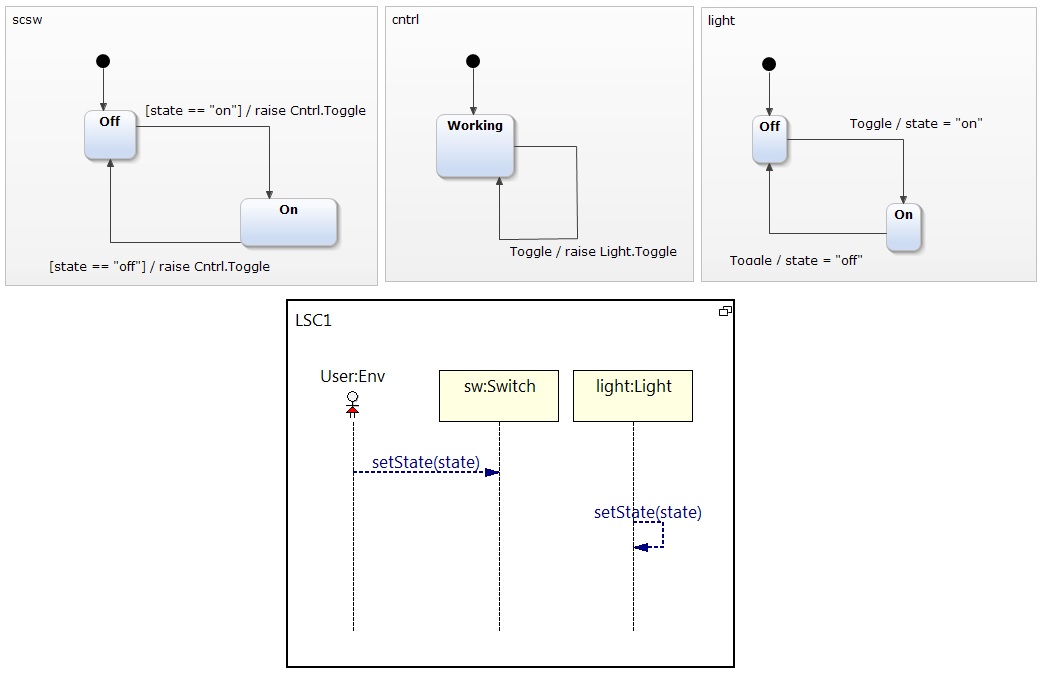}
\caption{The Switch-Light system: fully implemented}
\label{SL3}
\end{figure}

Gradually continuing with the implementation, the statechart of the controller can be added, and then the one for the light. Each time a statechart takes responsibility for the actual triggering of events,  the corresponding events in the LSCs are modified to be monitored, and can even be removed. Fig.~\ref{SL3} shows the statecharts of the three components, with the original requirement now in monitored mode. This model is actually the final implementation of this very small system, since all components are fully implemented. The LSC can be omitted at runtime, or it can run together with the statecharts,  in order to confirm at run time that the execution is indeed consistent with the requirements.

Although the example described here is extremely simple, it demonstrates  the main idea of our proposed development cycle, and the fact that although the implementation of the various components is incremental, the system can be executed in full at any time during the development cycle.

\section{Integration Semantics and Implementation} \label{Implementation}

\subsection{Semantics}
\begin{figure} [h!]
\centering
\includegraphics[scale=0.5]{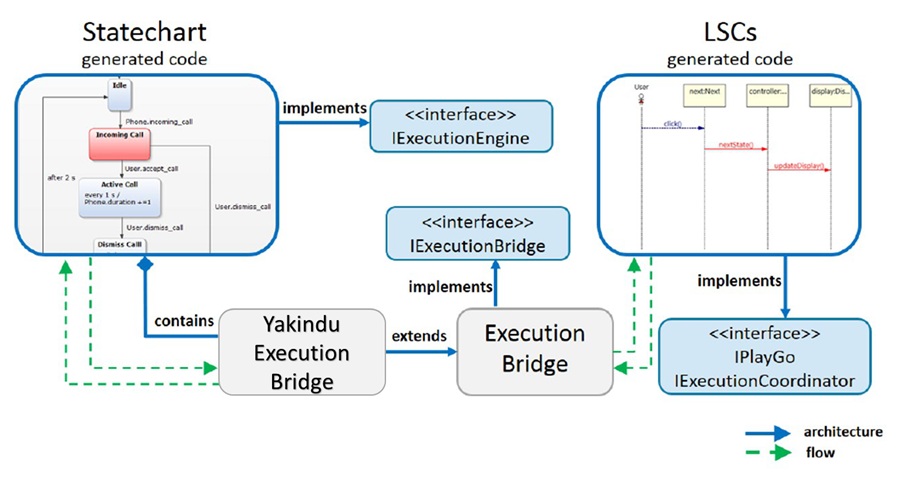}
\caption{High level architecture of our implementation}
\label{Architecture}
\end{figure}

Integrating any two run-time platforms calls for many decisions involving, e.g., mapping one's concept to the other's, concurrency and priority in execution, data sharing, messaging protocols, synchronization etc. 
The details of our implementation semantics are provide in the supplementary material, at \url{www.b-prog.org/sctlscSupp}~. 
Briefly, the key decisions that we made are as follows:  

\begin{enumerate}
\item{\textbf{PlayGo is the host environment,} controlling both development environments, with smooth switching between the two, and the runtime environment, with the coordinated execution, data sharing and message exchanges. The runtime architecture is depicted in Fig.~\ref{Architecture}. It relies on {\it Execution Bridge} to be able to interface with multiple kinds of models (Yakindu and others) and with multiple instances of any given model.}
\item{\textbf{The internal clocks of the two systems are synchronized.}} 
\item{\textbf{The generated Java code}  (of both PlayGo and Yakindu), can run without the development environments.}
\item{\textbf{Triggered Statecharts events have priority} over LSC events that are enabled and ready to execute at the same time. This choice 
stems from the desire to allow the implementation, which is commonly developed later, and must run,  to refine, and if needed, override the specification.}
\item{\textbf{Statecharts events that are forbidden in an LSC will nevertheless occur} and the resulting violation will be reported --- as opposed to deferring the event until is allowed, as would be done with forbidden LSC events. The rationale for this choice is the same as in the previous item.} 
\item{\textbf{The object model is shared} between the two platforms.}
\end{enumerate}

(In the future  we plan to make it possible for the user to choose semantic variations via plug-in code for event selection and execution-order policies.)  In addition to the above, we provide a detailed mapping between LSC events  (associated with messages or parameterized method calls) and the corresponding Statechart event names defined under object interfaces in Yakindu.

\subsection{Revisiting the railcar system} \label{Railcars}

\begin{figure} [h!]
\centering
\includegraphics[scale=0.45]{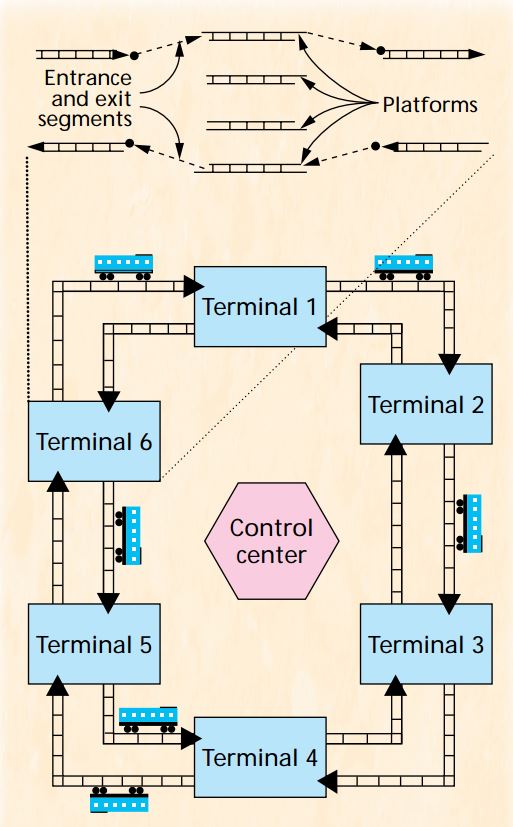}
\caption{The railcar system}
\label{RailcarUI}
\end{figure}

We  now proceed to  illustrate the capabilities of the methodology, and its semantics and the supporting tools, via the example appearing in the paper that introduced object-oriented statecharts \cite{HG97} 
(see Fig.~\ref{RailcarUI}), bringing the inter-object vs. intra-object duality to some kind of closure. 
For lack of space, our account here is rather brief, and a more detailed description appears in  the supplemental material at~\url{www.b-prog.org/sctlscSupp}~.

The setting is as follows. Multiple terminals are connected by a cyclic path consisting of two rail
tracks, one for each travel direction.
 Several railcars (abbr. cars hereafter) transport passengers
between terminals. A control center coordinates all activities.
Each terminal has multiple platforms, and the  incoming and  outgoing rail segments are each connected
to a short adjustable rail segment, within the terminal, which can be linked to any  of the
platforms.

Here now are some requirement scenarios.  They clearly illustrate the stand-alone, inter-object `story' nature of scenarios in general:
\begin{itemize}
\item {\it Car approaching terminal}. When the car is 100 yards from the terminal, the system allocates
a platform and an entrance segment, and, if the car is only passing through, also an exit segment. If the allocation is
 not completed when the car is within 80 yards from the terminal, the car must stop.
\item {\it Car departing terminal}. A car departs the terminal
 90 seconds after arrival. The system
connects the platform to the outgoing track
via the exit segment, engages the car’s engine, and
turns off the destination indicators on the terminal's
destination board. The car can then depart,
unless it is within 100 yards behind another car.
\item
{\it Passenger in terminal}. When a passenger is in a terminal
and no car in the terminal is traveling 
in the desired direction, the passenger can push a
destination button, and wait until a car arrives.
If the terminal contains an idle car, it is 
assigned to that destination, otherwise 
a car is sent in from another terminal. The
system indicates car availability on the destination board.
\end{itemize}

\begin{figure} [h!]
\centering
\includegraphics[scale=0.6]{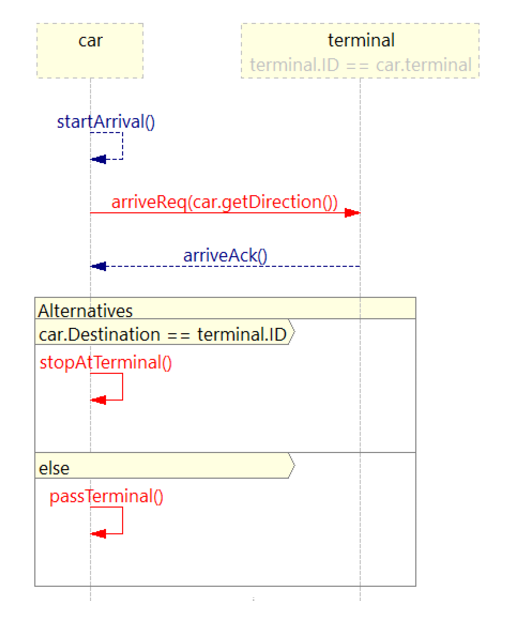}
\includegraphics[scale=0.5]{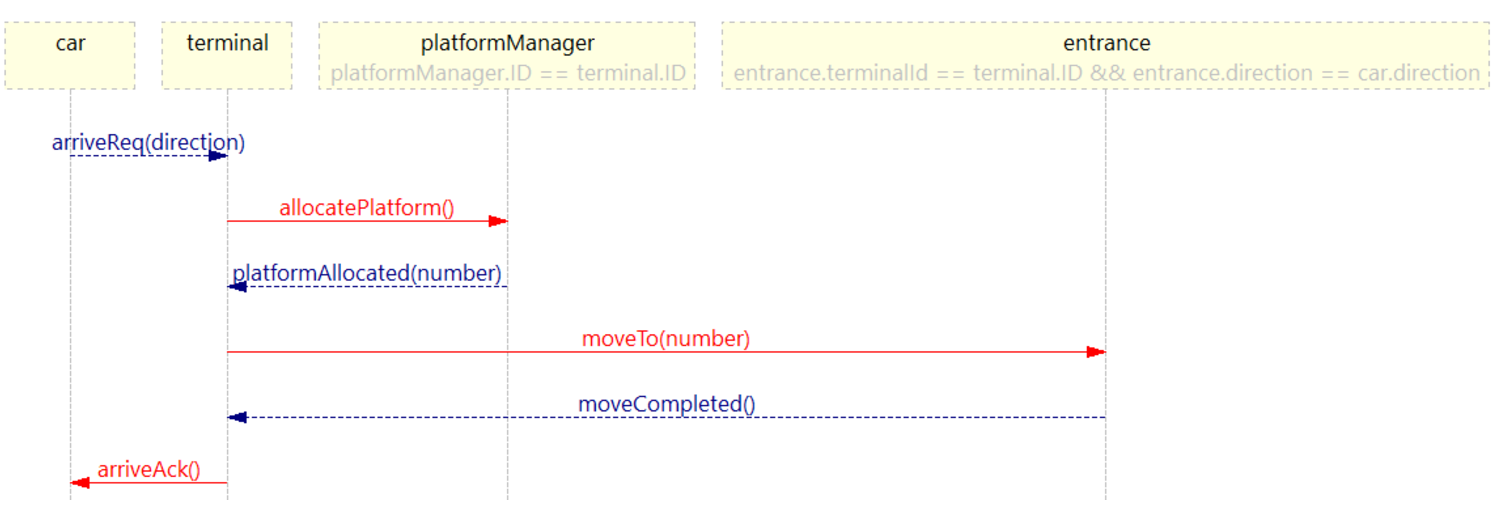}
\caption{Top LSC: Car Arrival at a Terminal.  The car first calls {\it startArrival}. It then sends an {\it arriveReq} message to that terminal and  waits for acknowledgment. Depending on whether the next terminal is its destination or not,  it stops or  passes through. ~~~Bottom LSC:  Arrival Request. The terminal asks the platform manager to allocate a platform and waits for an approval containing the allocated platform's number. Then the car is sent to the corresponding entrance.
 Note in both LSCs the symbolic lifelines, which are  concretized (instantiated) to specific objects via a binding expression; for example, in the top LSC, the car's {\it terminal} property is compared with the ids of all terminals. }
\label{RailCarLSCa}
\end{figure}

\begin{figure} [h!]
\centering
\includegraphics[scale=0.35]{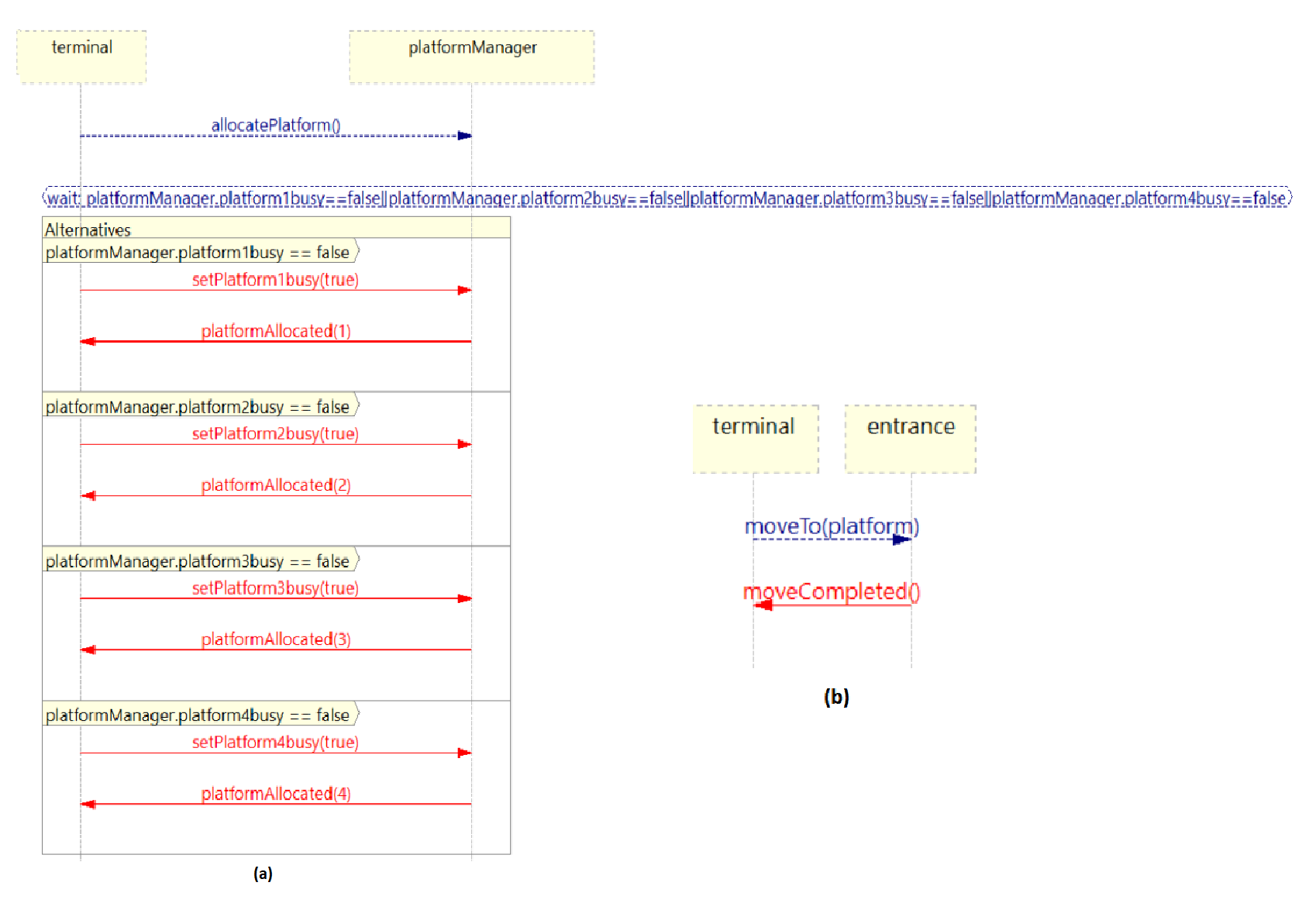}
\caption{(a) Platform Allocation: : the platform manager waits for some platform to become available, allocates the first available platform and marks it as busy~~~(b) A simple LSC handling the handshake between the terminal and the entrance.
}
\label{RailCarLSCcd}
\end{figure}

 In~\cite{HG97}, this system was programmed using object-oriented Statecharts; see Fig.~\ref{statechartexample} for one of those. In the new approach, we start by formalizing the requirements as LSCs, as exemplified in  Figs.~\ref{RailCarLSCa} and ~\ref{RailCarLSCcd}. 

\begin{figure} [h!]
\centering
\includegraphics[scale=0.55]{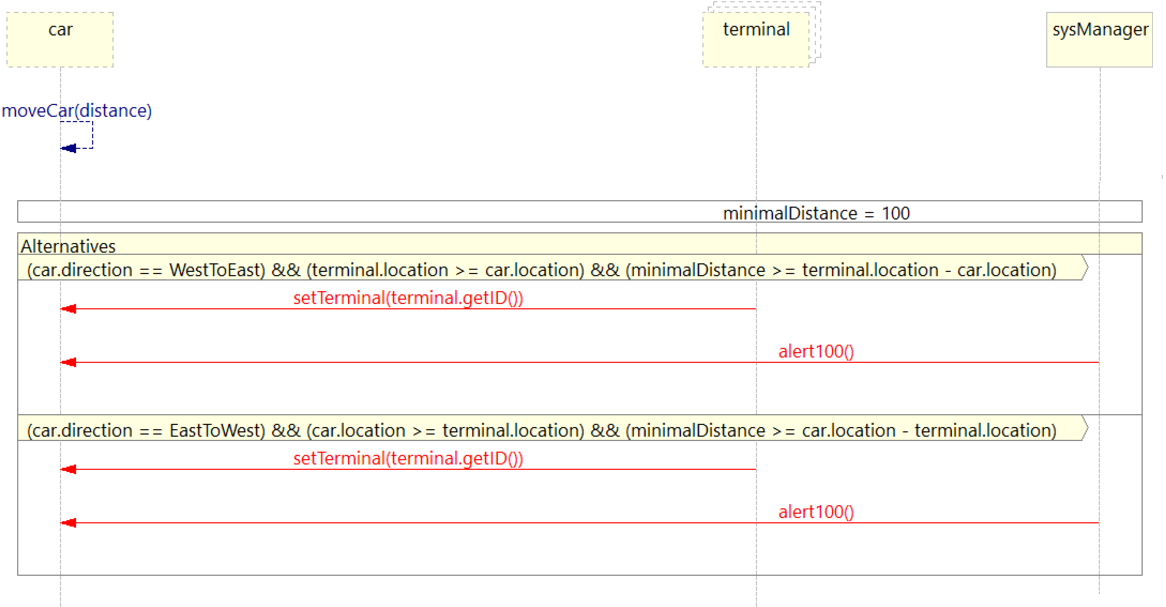}
\caption{Alert 100}
\label{RailCarLSCe}
\end{figure}

The LSC in Fig.~\ref{RailCarLSCe} specifies that every time a car moves, each terminal checks whether the car is moving in its direction and has passed the minimal distance. Note the use of symbolic lifelines with multiplicity, indicating a scenario that applies to multiple terminals. Clearly, the implementation will differ from what is described in the scenario, but since we are in the requirements phase, we keep our scenarios abstract and ignore implementation and efficiency issues. If there is a terminal that meets the conditions of direction and proximity, it informs the car that it is approaching and sends its specific terminal number (following this action, the car sets its terminal variable to this number) and then the system manager sends an {\it alert100} signal to the car.

\begin{figure} [h!]
\centering
\includegraphics[scale=0.7]{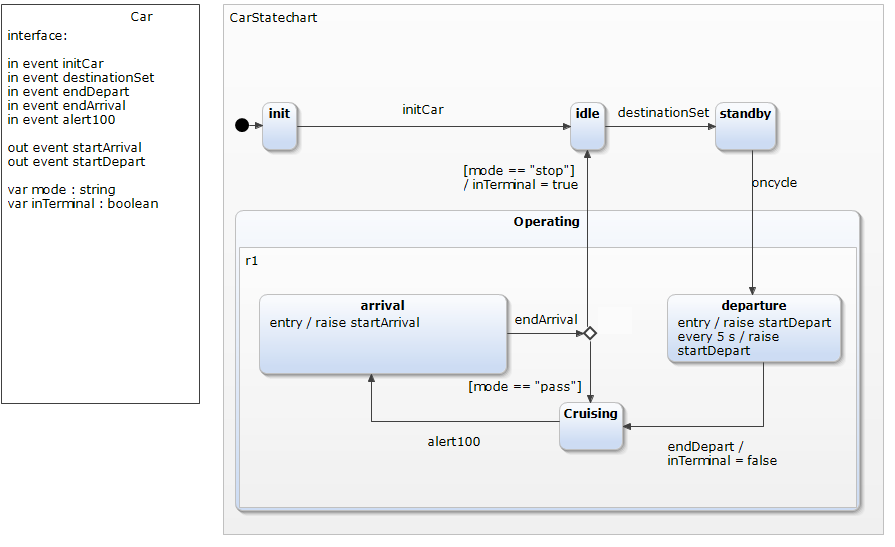}
\caption{The car statechart}
\label{CarStatechart}
\end{figure}

In our implementation, we chose to implement the car at the intra-object level, while leaving the other objects (the car's `environment') at the requirements inter-object level. 
Therefore, the car's statechart (Fig.~\ref{CarStatechart}) reacts to all the input signals that are sent to the car and raises the output signals expected by other objects.

\begin{figure} [h!]
\centering
\includegraphics[scale=0.55]{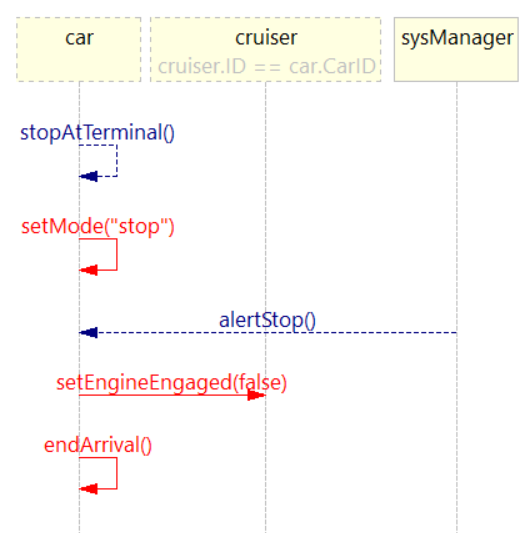}
\caption{Stop at terminal}
\label{RailCarLSCf}
\end{figure}

The car's statechart is quite straightforward, so we will  focus on the interaction with the LSCs. When the car is in state {\it cruising}, it waits for the {\it alert100} signal and reacts by moving to the {\it arrival} state and raising the {\it startArrival} signal (event). This signal belongs to the {\it Car} default interface and is therefore handled in the LSC as a self method call. 
This event triggers the LSC in figure~\ref{RailCarLSCa}. The car then waits for the {\it endArrival} event and moves to the {\it idle} or {\it cruising} state, depending on whether or not it should stop at that terminal. The {\it endArrival} event is raised  by the LSC that handles the car's passing through the terminal and the LSC that handles its stopping at the terminal (Fig. \ref{RailCarLSCf}). 

\section{Related Work} \label{RelatedWork}

Transitions from scenario-based specifications to code via state machines can be seen in the large amount of work 
on synthesizing finite automata from MSC, sequence diagrams and LSCs (see, e.g., ~\cite{liang2006comparative,Segall2012Synth, Greenyer2016a} and references therein).  

%
However, synthesis is often impractical, as 
the size of the resulting composite automaton grows exponentially with the number of orthogonal scenarios and the allowed range of variable data and event parameters.  
To help combat state explosion, synthesis solutions often constrain the supported scope of certain expressive features in the original scenario language. 

Execution of final systems via the play-out of executable, scenario-based specifications (as envisioned, e.g., in~\cite{DHpubs} paper 174),  is not    
ready yet to fully materialize. More robust solutions are needed for dealing with legacy code and with engineers' preferences of languages. We also need  adequate approaches for decomposing distributed systems.
Hence, we believe  that there is a need for the kind of integration mechanism and methodology proposed here, which allow human engineers to conduct a well-controlled gradual transitioning from requirements to system implementation.

In~\cite{damm2002statemateLSC}, the authors discuss showing different system views at various abstraction levels,
verifying statecharts against LSC specifications and using  LSCs to generate test vectors for the statecharts.
However, the LSCs and statechart models in~\cite{damm2002statemateLSC} are separate and their execution is not directly integrated.
>>> NEED REFS FOR RHAPOSODY AND INTERPLAY IN THIS PARAGRAPH <<<<<<
The  Rhapsody tool supports monitoring 
statechart execution against the sequence diagrams specifications, but the sequence diagrams cannot influence the execution. 
InterPlay is a tool developed in our group in order to link Statecharts and LSC. It essentially provides a gateway that propagates and translates events between independently running LSCs and statechart engines. 
The contributions of the present research over Interplay include: a fully-shared object model between the LSCs and the statecharts; an integrated and synchronized execution semantics, and a supporting mechanism that can also operate without the presence of the (heavy) development environments (PlayGo and Yakindu in this case).

In recent years, a number of efforts have been made to enable the joint
simulation and analysis of models developed in different formalisms. These include  
Ptolemy II, with its multiple models of computation (MoC);  
ModHel'X, which combines semantics of multiple languages; the Epsilon Merging Language (EML),
which provides a rule-based language for merging models of diverse
meta-models and technologies; Reusable Aspect Models (RAM), which integrates structural
models, message views and state views using an aspect-oriented modeling technique; and, the GEMOC- based BCOoL coordination language  which allows the specification of diverse semantics and integration between multiple languages.  We have not been able to find a system development environment where the \emph{execution} (or simulation) of 
(LSC-like) multi-modal scenarios and (Statechart-like) state machines can be truly integrated, with
well-defined semantics.

In separate but related work, we have amalgamated Statecharts with SBP, by extending Yakindu to allow associating individual states with requested and blocked events, and then enhancing the Yakindu event triggering mechanism  to deal with such specifications~\cite{morse18SCSBP}. While this development allows an engineer to specify both scenarios and state-based reactivity in a single formalism, it is yet to be investigated how the intuitiveness of the scenario's `story', and the clarity of the roles played by the participating objects, which are  key tenets of sequence diagrams and LSCs,  can be accomplished in Statecharts. Is this an issue of design patterns, or of visual formatting? Or is this yet another issue altogether?

\section{Conclusion and Future Work}
\label{Conclusion}

We have presented a development environment and a methodology for incremental system development, starting with intuitive requirement scenarios and ending with object-oriented state machines, where throughout the process all artifacts are analyzable and executable, enabling simulation and validation at all stages. In addition, the availability of powerful versions of the two modeling approaches implemented in a single integrated tool, simplifies developers' choice of the most suitable and naturally-fitting language for the various parts of  the system.

Future developments of the integration include: (1) a more straightforward mapping between parametrized LSC events and Statechart events; 
(2) enabling semantic variations via user-supplied code ; (3) enabling integration also with components written in standard procedural languages ; (4) incorporating into the integrated platform research results that are available in either SBP and Statecharts, such as formal verification, context-awareness, natural language input, execution with look-ahead (\emph{smart play-out}), run-time learning, and more (see, e.g., in~\cite{DHpubs}, publications 230, 190, 112, 217). 

We believe that a single tool and methodology
for developing executable models in both inter- and intra-object approaches --- supporting both 
requirement specification  and implementation phases, and with means for smooth and semantically consistent transition between the two --- 
can have a dramatic impact on the cost and quality of complex systems development. 

\section*{Acknowledgments}
This work has been supported in part by a grant to David Harel
 from the Israel Science Foundation,
 the William Sussman Professorial Chair of Mathematics,
 and the Estate of Emile Mimran. 

\bibliographystyle{abbrv}
\bibliography{GlobalBib3SCTLSC}

\noindent  \textbf{David Harel:}: Prof. David Harel is the Vice President of the Israel Academy of Sciences and Humanities, and has been at the Weizmann Institute of Science since 1980, serving in the past as Dean of its Faculty of Mathematics and Computer Science. He has worked in logic and computability, software and systems engineering, modeling biological systems and more. He invented Statecharts and co-invented Live Sequence Charts. Among his books are “Algorithmics: The Spirit of Computing” and “Computers Ltd.: What They Really Can't Do”. His awards include the ACM Karlstrom Outstanding Educator Award, the Israel Prize, the ACM Software System Award, the Eme”t Prize, and five honorary degrees. He is a Fellow of ACM, IEEE and AAAS, a member of the Academia Europaea and the Israel Academy of Sciences, and a foreign member of the US National Academy of Engineering and the American Academy of Arts and Sciences.

\noindent \textbf{Rami Marelly:} Dr. Rami Marelly holds a Ph.D. in computer science from the Weizmann Institute of Science. His research was about specifying and executing behavioral requirements using the Play-in/Play-out approach. Rami held key technology leadership positions in the Israeli Air Force including Head of C4I Systems Dept. and Head of Aerial ISR Systems. As head engineer, Rami led the IAF IT transformation towards network centric warfare and was responsible for the development of networking, avionics, simulators, C4I and security systems. After retiring (Col. res.) from the IAF, Rami co-founded Cue, a consulting firm. He teaches advanced academic courses in systems engineering and volunteers as a mentor in various FIRST robotics projects

\noindent \textbf{Assaf Marron:} Dr. Assaf Marron is a researcher at the Weizmann Institute of Science Computer Science and Applied Mathematics Department.  His research interests include software engineering  and artificial intelligence. He holds a PhD in computer science from the University of Houston. Prior to joining the Weizmann Institute, he held senior management and technical positions in leading companies including IBM and BMC Software. He is the inventor or co-inventor of several patents.

\noindent \textbf {Smadar Szekely} Mrs. Smadar Szekely was the head of software team in the research group of Prof. David Harel, between 2009 and 2018. Prior to this she worked as R\&D Director at Sungard Corp. where she created complex mission-critical software products and defined corporate software development processes. 
Smadar holds a B.A. in Computer Science from Tel-Aviv Yaffo College.
\end{document}